\documentstyle[12pt]{article}
\author{V.I.Nazaruk\\
Institute for Nuclear Research of RAS, 60th October\\
Anniversary Prospect 7a, 117312 Moscow, Russia\\
E-mail: nazaruk@al20.inr.troitsk.ru}

\title{Cause of Disparity between the Results for the $n\bar{n}$
       Transition Problem}
\date{}
\begin{document}

\maketitle
\begin{abstract}

We show that there is a double counting in the standard model of $n\bar{n}$
mixing in the medium, resulting in full cancellation of leading terms. The
direct calculation of $n\bar{n}$ transition, annihilation is performed. For
lower limit on the free-space $n\bar{n}$ oscillation time we get  $\tau_{min}
\sim 10^{31}y$.
\end{abstract}

\vspace{5mm}

\newpage

\setcounter{equation}{0}

Any information on the occurrence of $n\bar{n}$ oscillation[1,2] is important
in order to discriminate among various grand unified theories. The most direct
limit on the free-space $n\bar{n}$ oscillation time $\tau_{n\bar{n}}$
is obtained using free neutrons:\ $\tau_{n\bar{n}}>10^{7}s$[3]. Alternatively,
a limit can be extracted from the nuclear annihilation lifetime $T$ measured
in proton-decay type experiments:\ $\tau_{n\bar{n}}>10^{8}s\sim 1y$ (see,
for example, Ref.[4]). The calculations involved were based on the potential
model of $n\bar{n}$ mixing in the medium. In this letter the model independent
approach is presented.

{\bf 1.}In Ref.[5] for free-space $n\bar{n}$ oscillation time the limit $\tau_
{min}=3\cdot10^{31}y$ was obtained, which increases the previous one (see, for
example, Ref.[4]) by 31 orders of magnitude. First off all we expose a
drawback hidden in the standard model. In the standard approach (labelled
bellow as potential model) the $n\bar{n}$ transitions in a medium are
described by Schrodinger equations
\begin{equation}
\label{1}(i\partial_t+\nabla^2/2m-U_n)n(x)=\epsilon \bar{n}(x),\;\;\;\;
(i\partial_t+\nabla^2/2m-U_{\bar{n}})\bar{n}(x)=\epsilon n(x).
\end{equation}
Here $\epsilon=1/\tau_{n\bar{n}}$ is a small parameter[4]; $U_n$ and
$U_{\bar{n}}$ are the self-consistent neutron potential and $\bar{n}$-nucleus
optical potential respectively. For $U_n=const.$ and $U_{\bar{n}}=const.$ in
the lowest order on $\epsilon$ the probability of the process is
\begin{equation}
\label{2}W_{pot}(t)=1-\mid U_{ii}(t)\mid ^2=2ImT_{ii}(t),\;\;\;\;T_{ii}(t)=
i(\epsilon/\delta U)^2[1-i\delta Ut-\exp (-i\delta Ut)],
\end{equation}
where $U=1+iT,T$ are the evolution and $T$-operator respectively;
\begin{equation}
\label{3}\delta U=U_{\bar{n}}-U_n,\;\;\;U_{\bar{n}}=ReU_{\bar{n}}-i\Gamma /2,
\end{equation}
$\Gamma\sim 100 MeV$ is the annihilation width of $\bar{n}$-nucleus state.

What is meant by $W_{pot}(t)$? Let us take the imaginary part of Eq.(16) of
Ref.[5]
\begin{equation}
\label{4}2ImT_{ii}(t)=\epsilon^2t^2-\epsilon^2\int_0^tdt_{\alpha}\int_0^
{t_{\alpha}}dt_{\beta}2ImT_{ii}^{\bar{n}}(\tau ),
\end{equation}
$\tau=t_{\alpha}-t_{\beta}$. Here $T(t),T^{\bar{n}}(\tau )$ are the $T$-matrix
of the whole process and $\bar{n}$-nucleus decay respectively.  From the
condition of probability conservation $\sum_{f}\mid U_{fi}\mid ^2=1$ we have
\begin{equation}
\label{5}2ImT_{ii}=\sum_{f\neq i}\mid T_{fi}\mid ^2+\mid T_{ii}\mid ^2,\;\;\;
\;\;\sum_{f\neq i}\mid T_{fi}(t)\mid ^2=W(t).
\end{equation}
The process probability $W(t)$ will be specified bellow.  $\mid T_{fi}\mid ^2
\sim \epsilon^2$, whereas $\mid T_{ii}\mid ^2\sim\epsilon^4$. So for the
l.h.s. of Eq.(4) $2ImT_{ii}(t)=W(t)$, that was taken into account in (2). For
the $T$-matrix of annihilation nucleus decay $T_{ii}^{\bar{n}}(\tau )$ Eq.(5)
has the form
\begin{equation}
\label{6}2ImT_{ii}^{\bar{n}}(\tau)=\sum_{f\neq i}\mid T_{fi}^{\bar{n}}
(\tau )\mid ^2+\mid T_{ii}^{\bar{n}}(\tau )\mid ^2.
\end{equation}
The annihilation nucleus decay is nonperturbative process and $\mid T_{ii}^
{\bar{n}}\mid ^2\sim\sum_{f\neq i}\mid T_{fi}^{\bar{n}}\mid ^2 $.
Now Eq.(4) has the form
\begin{equation}
\label{7}W(t)=\epsilon^2t^2-\epsilon^2\int_0^tdt_{\alpha}\int_0^
{t_{\alpha}}dt_{\beta}\mid T_{ii}^{\bar{n}}(\tau )\mid ^2-\epsilon^2
\int_0^tdt_{\alpha}\int_0^{t_{\alpha}}dt_{\beta}\sum_{f\neq i}\mid T_{fi}^
{\bar{n}}(\tau )\mid ^2
\end{equation}
Let us calculate $T_{ii}^{\bar{n}}$ and $T_{fi}^{\bar{n}}$ in the framework
of potential model. The wave function of initial state obeys equation
\begin{equation}
\label{8}i\frac{\partial \Phi }{\partial t}=H_0\Phi,\;\;\;H_0=-\nabla^2/2m+U_n.
\end{equation}
In $t=0$ the interaction $\delta U$ is turned on. We have
\begin{equation}
\label{9}i\frac{\partial \Psi }{\partial t}=(H_0+\delta U)\Psi ,\;\;\;
\Psi(0)=\Phi (0).
\end{equation}
The projection to the initial state and $T$-matrix at $t=\tau $ are
\begin{equation}
\label{10}<\Phi \mid \Psi >=U_{ii}^{\bar{n}}(\tau)=\exp (-i\delta U\tau),
\end{equation}
\begin{equation}
\label{11}T_{ii}^{\bar{n}}(\tau )=i[1-\exp (-i\delta U\tau )],\;\;\;\;\;
\sum_{f\neq i}\mid T_{fi}^{\bar{n}}(\tau )\mid ^2=1-\mid U_{ii}^{\bar{n}}
(\tau)\mid ^2=1-e^{-\Gamma\tau}=W_{\bar{n}}(\tau),
\end{equation}
where $W_{\bar{n}}(\tau)$ is the $\bar{n}$-nucleus decay probability. Note that
$\Gamma $ corresponds to all $\bar{n}$-nucleus interactions followed by
annihilation. However, the main contribution gives the annihilation without
rescattering of $\bar{n}$[6], because $\sigma _{ann}>2\sigma _{sc}$.
Substituting these expressions in (7), one obtains the potential model result
(2).

Therefore, the finite time approach was verified by the example of exactly
solvable potential model. It is involved in Eq.(4) as a special case.

Let us return to Eq.(7). It is at least unclear. 1.The first term is
free-space $n\bar{n}$ transition probability. Matrix elements $T_{ii}^{\bar
{n}}$ and $T_{f\neq i}^{\bar{n}}$ describe transitions $(\bar{n}-nucleus)
\rightarrow (\bar{n}-nucleus)$ and $(\bar{n}-nucleus)\rightarrow
(annihilation\; products)$ respectively. So the first and second terms
correspond to $\bar{n}$-nucleus in the final states. However, in the
experiment only annihilation products are detected and the result should be
expressed as $T_{f\neq i}^{\bar{n}}$ solely. Moreover, $\bar{n}$-nucleus
decays into final state products identical with those given by third term.
This suggests that potential model contains the double counting. Expression
$1-\mid U_{ii}\mid ^2$ from Eq.(2) describes the inclusive decay of initial
state and so the $n\bar{n}$ transition with $\bar{n}$-nucleus in the final
state is also included in $W_{pot}$, unless additional limits are imposed. To
obviate the double counting the annihilation products in the final state
should be fixed. 2.Let us $\mid \delta Ut\mid \ll 1$. (This regime occurs in
other problems.) When $\Gamma =0$, the third term equals to zero. When $\Gamma
\neq 0$, the contribution of the third term is negative and $dW/d\Gamma <0$,
whereas the opening of the new channel (annihilation) should increase $W$.

How much is the probable error? Contributions of the second and third terms
are: $x_2=-\epsilon^2t^2/2+F_2$, $x_3=-\epsilon^2t^2/2+F_3$. Functions $F_{2,
3}$ contain the terms proportional to $t$ and $\exp (-i\delta Ut)$. So the
$\epsilon^2t^2$ term produced by third term is fully canceled. This is a
consequence of double counting. Therein lies a reason of discrepancy between
ours and potential model results. Solving Eqs.(1) by method of Green functions
we will reach the same results. We have started from Eq.(4) only for
verification of finite time approach.

As noted in[5], Eqs.(11) and (2) can be also obtained by means of microscopic
variant of potential model (zero angle rescattering diagrams of $\bar{n}$). In
this case the Hamiltonian of $\bar{n}$-medium interaction is $H=\delta U$. The
same calculation was repeated by Dover et al.[4]. They substitute $H=-i\Gamma
/2$ in (4) and obtaine (2). On the basis of this and only this they refute the
result of Ref.[5]. In other words they refute our limit because it differs
from prediction of potential model ($H=-i\Gamma /2$[4]).

Our concern is with $\sum_{f\neq i}\mid T_{fi}\mid ^2$. It is connected with
diagonal matrix element by Eq.(5):
\begin{equation}
\label{12}2ImT_{ii}=\sum_{f\neq i}T_{if}^*T_{fi}.
\end{equation}
Calculation of $T_{ii}$ is determined by r.h.s., namely, the cut corresponding
to $T_{ii}$ must contain only annihilation products, that is not in accordance
with Eq.(7). It includes superfluous "incorrect" states $f'=(\bar{n}-nucleus)$
. Note, that eigenfunctions of $H_0+\delta U$ do not form the complete
orthogonal set. Because of this the $(\bar{n}-nucleus)$ (described by $U_{\bar
{n}}$) also may not appear in Eq.(12) as intermediate state. So model (1) is
inapplicable to the problem under study because it automatically leads to
incorrect matrix element $T_{ii}$. Elimination of superfluous trajectories
from $T_{ii}$ means the direct calculation of $T_{fi}$. We can write $W=2ImT_
{ii}$, however $T_{ii}$ should be calculated by means of r.h.s. of Eq.(12).

{\bf 2.}In Ref.[5] the first and third terms were taken into account. The
second one was omitted. The first term reproduces low density limit and has a
sense for $n\bar{n}$ transitions in the gas. This scheme is not quite correct.
In this paper we present the direct calculation of the process $(nucleus)
\rightarrow (\bar{n}-nucleus) \rightarrow (annihilation\; products)$. We have
\begin{equation}
\label{13}
\begin{array}{c}
<f\mid U(t,0)-I\mid0n_p>=iT_{fi}(t)=\\
\sum_{k=1}^{\infty}(-i)^{k+1}<f\mid \int_0^tdt_1...\int_{0}^{t_{k-1}}dt_k
\int_{0}^{t_k}dt_{\beta }H(t_1)...H(t_k)H_{n\bar{n}}
(t_{\beta })\mid0n_p>,
\end{array}
\end{equation}
where
\begin{equation}
\label{14}H(t)=(all\;\; \bar{n}-medium\;\; interactions) - U_n,\;\;\;\;
H_{n\bar{n}}(t)=\epsilon \int d^3x(\bar{\Psi }_{\bar{n}}\Psi _n+h.c.),
\end{equation}
$H+H_{n\bar{n}}=H_I$. Here $\mid \!0n_p\!>$ is the state of medium containing
the neutron with 4-momenta $p=({\bf p}_n^2/2m+U_n,{\bf p}_n)$, $<\!f\!\mid$ is
the annihilation products; $H_{n\bar{n}}$ is the oscillation Hamiltonian[4].
In the case of the formulation of the $S$-matrix problem $(t,0)\rightarrow
(\infty ,-\infty)$ Eq.(13) in the momentum representation includes the
singular propagator $G=1/(\epsilon_n-{\bf p}^2_n/2m-U_n)\sim 1/0$. Taking into
account that $H_{n\bar{n}}\mid\!0n_p\!>=\epsilon \mid\!0\bar{n}_p\!>$, we
change the integrating order and obtain
\begin{equation}
\label{15}
\begin{array}{c}
T_{fi}(t)=- \epsilon \int_{0}^{t}dt_{\beta }iT_{fi}^{\bar{n}}(t-t_{\beta }),\\
iT_{fi}^{\bar{n}}(\tau )=\sum_{k=1}^{\infty}(-i)^k\int_
{t_{\beta }}^{t}dt_1...\int_{t_{\beta }}^{t_{k-1}}dt_k<f\mid H(t_1)...H(t_k)
\mid0\bar{n}_p>,
\end{array}
\end{equation}
where $T_{fi}^{\bar{n}}$ is an exact amplitudes of $\bar{n}$-nucleus decay,
$\mid\!0\bar{n}_p\!>$ is the state of medium containing the $\bar{n}$ with
4-momenta $p$; $\tau=t-t_{\beta}$. 4-momenta of $n$ and $\bar{n}$ are equal.

The 2-step process was reduced to the annihilation decay of $\bar{n}$-nucleus.
(The slightly different method is that antineutron Green function is separated
[5].) It is seen from (13), and (15) that both pre- and post- $n\bar{n}$
conversion spatial wave function of the system coincide
\begin{equation}
\label{16}\mid\!0n_p\!>_{sp}=\mid\!0\bar{n}_p\!>_{sp}.
\end{equation}
We would like to stress that in potential model the picture of $\bar{n}$-
nucleus formation is precisely the same. Really, let us the $n\bar{n}$
conversion takes place at $t=0$. Solution of Eqs.(1) is continuous and $\Psi
(-0)=n=\Psi(+0)=\bar{n}$, that is identical to (16). (See also Eqs.(8),(9).)
Note that Eq.(4) was obtained in perfect analogy to (15). In particular, for
$T_{ii}$ and $T_{ii}^{\bar{n}}$ condition (16) was fulfilled. Hereafter, the
potential model of $\bar{n}$-medium interaction (block $T^{\bar{n}}$) was used
and $W_{pot}$ was reproduced, which also corroborates the picture of $\bar{n}$-
nucleus formation given above.

In both models the stage of $n\bar{n}$ conversion is identical. The basic
difference centers on the next stage - annihilation. In the potential model
$T_{ii}^{\bar{n}}$ is calculated end used in Eq.(7), which is wrong. We
calculate $T_{fi}^{\bar{n}}$ starting from the same point (16). The result
will be expressed through $\Gamma $ (see Eqs.(19),(11)), but not $\delta U$,
as it usually is in calculation of decays.

Let us take $t=T=6.5\cdot10^{31}y$[7], where $T$ is nuclear annihilation
lifetime. The characteristic annihilation time of $\bar{n}$ in nuclear matter
is $1/\Gamma \sim 10^{-24}$s. When $\tau \gg 1/\Gamma$, $T_{fi}^{\bar{n}}(\tau
)$ reaches its asimptotic value $T_{fi}^{\bar{n}}$:
\begin{equation}
\label{17}T_{fi}^{\bar{n}}(\tau \gg 1/\Gamma )=T_{fi}^{\bar{n}}(\infty )=
T_{fi}^{\bar{n}}=const.
\end{equation}
The expressions of this type are the basis for all $S$-matrix calculations.
(Measurement of any process corresponds to some interval $\tau $. Consequently,
it is necessary to calculate $U(\tau )$. Replacement $U(\tau )\rightarrow S(
\infty )$ is equivalent to (17).) From (15) and (17) we have
\begin{equation}
\label{18}T_{fi}(T)=-i\epsilon [\int_{0}^{T-1/\Gamma }dt_{\beta }T_{fi}^{\bar
{n}}(T-t_{\beta })+\int_{T-1/\Gamma }^{T}dt_{\beta }T_{fi}^{\bar{n}}(T-t_
{\beta })]\sim -i\epsilon TT_{fi}^{\bar{n}}.
\end{equation}
Really, $\mid T_{fi}^{\bar{n}}(\tau)\mid \leq 1$ because $W_{fi}(\tau )=\mid
T_{fi}^{\bar{n}}(\tau)\mid ^2\leq 1$; $T=2\cdot10^{39}$s, or $T\sim 10^{63}/
\Gamma $ in units of $1/\Gamma $. Obviously, the contribution of second term
is negligible. The probability of the whole process is
\begin{equation}
\label{19}W(T)=\sum_{f\neq i}\mid T_{fi}(T)\mid ^2\sim \epsilon^2T^2
 \sum_{f\neq i}\mid T_{fi}^{\bar{n}}(T)\mid ^2\sim \epsilon^2T^2,
\end{equation}
where Eq.(11) have been taken into account. The limit for $\tau_{n\bar{n}}$ is
obtained from the inequality $W(T)<1$. For $T=6.5\cdot10^{31}y$[7] we have
$\tau_{min}\sim 10^{31}y$.

{\bf 3}.Let us return to the reason of enormous quantitative disagreement
between the our and potential model results. The strong result sensitivity was
to be expected. Really, the $S$-matrix amplitude $M_s$, corresponding to
$n\bar{n}$ transition, annihilation diverges:
\begin{equation}
\label{20}M_s=\epsilon\frac{1}{\epsilon_n-{\bf p}^2_n/2m-U_n}M\sim \frac{1}{0},
\end{equation}
where $M$ is the annihilation amplitude. This is infrared singularities
conditioned by zero momentum transfer in the $\epsilon $-vertex. It is easy to
understand that $M_s\sim 1/0$ for any bound state wave function of neutron
(i.e., for any nuclear model). On the other hand from Eqs.(1) it is clear that
in the potential model the energy is not conserved and becomes complex in the
$\epsilon $-vertex $M_A\rightarrow M_A+\delta U$ ($M_A$ is the nuclear mass).
The corresponding antineutron Green function is $G=1/(\epsilon_n-{\bf p}^2_n
/2m-U_{\bar{n}})=1/\delta U$. $\delta U=0$ is the peculiar point of $M_s$. So
$M_s$ is extremely sensitive to $\delta U$. (Usually, the $\delta U$-
dependence of $G$ is masked by momentum transferred $q$: $G^{-1}=(\epsilon_n-
q_0)-({\bf p}_n-{\bf q})^2/2m-U_n-\delta U$. We deal with 2-tail and $q=0$.)

Comparing (20) with (18) one sees that primitively the limit $\delta U
\rightarrow 0$ corresponds to replacement
\begin{equation}
\label{21}1/\delta U\rightarrow T.
\end{equation}
Certainly, we do not set $\delta U=0$, because $S$-matrix amplitude is not
considered at all. In the calculation of Eq.(13) the multiplier $T$ (see Eq.
(18)) arises automatically instead of $1/\Delta q$ in the $S$-matrix theory.
When $q\neq 0$ in the $\epsilon$-vertex, Eq.(13) leads to usual $S$-matrix
result (see below). Formal reason for the differences in the results is the
full cancellation of terms $\sim t^2$ in Eq.(7). Erroneous structure of (7)
conditioned by nonperturbative and 2-step character of the process. $q=0$
extremely reinforces the disagreement.

{\bf 4}.One additional comment is necessary regarding t-dependence of the whole
process probability $W(t)$. Eq.(19) has been obtained in the lowest
order on $\epsilon$. The precise distribution $W_{pr}(t)$ which allows for
the all orders on $\epsilon$ is unknown. However, $W$ is the first term of the
expansion of $W_{pr}$ and we can restrict ourselves to a lowest order
$W_{pr}=W$, as it usually is for rare decays. $W_{pot}$ is also calculated
in the lowest order on $\epsilon$.

The protons must be in very early stage of the decay process. Thus the
realistic possibility is considered[8-11] that the proton has not yet entered
the exponential stage of its decay but is, instead, subject to non-exponential
behavior which is rigorously demanded by quantum theory for sufficiently early
times. At first sight, since $\tau_{n\bar{n}}>10^{31}y$ for $n\bar{n}$-mixing
in nuclear the similar picture should be expected. In fact the situation is
more serious. We deal with two-step process. In attempting to calculate
$M_s$ and $\Gamma _s \sim \mid M_s\mid ^2$ in the framework of standard
S-matrix theory we get $\Gamma _s\sim 1/0$. So there is no sense to speak
about decay law $\exp (-\Gamma _st)$. It is necessary to calculate the
distribution $W(t)$ as it was done above.

Finally, we will touch upon the main points of Krivoruchenko's preprint[12].
(1) The $n\bar{n}$ transition, annihilation (two-step nuclear decay) and
particles motion in the classical fields are the different problems.
Describing the first one by Eqs.(1) we understand that this is an effective
procedure. From formal standpoint in the first and second cases the potentials
are complex and real respectively. Unfortunately, sometimes the literal
analogy between these problems is drawn[12]. (2) The initial Eq.(11) of Ref.
[12] must describe the $n\bar{n}$ transition, annihilation. However, the l.h.
s. of Eq.(11) is free of $\bar{n}$ -nucleus interaction at all. The r.h.s.
contains annihilation width $\Gamma$ (we stress this point) and coincides with
potential model result. We would like also to get the result without
calculation, but some difficulties emerge in reaching this goal.

We attempt to calculate the process amplitude starting from (14). The $S$-
matrix theory gives (20). The approach with finite time interval is infrared-
free. Its verification for diagrams with $q=0$ was made above by the example
of potential model. For nonsingular diagrams the test is obvious. Let us $q
\neq 0$ in the $\epsilon$-vertex. Appropriate calculation with finite time
interval (adiabatic hypothesis should be used) converts to the $S$-matrix
result: $T_{fi}=i\epsilon '(1/\Delta q){\cal T}^{\bar{n}}_{fi}$, where ${\cal
T}^{\bar{n}}_{fi}$ is the $S$-matrix amplitude of annihilation of virtual
$\bar{n}$ with 4-momenta $k=p-q$. Comparing with (18) one sees that limit
$\Delta q\rightarrow 0$ corresponds to replacement $1/\Delta q\rightarrow t$
(compare with (21)). Similar problem for matrix element $T_{ii}$ was solved in
Ref.[13]. Note, however, that there are essential differences between above
mentioned problems. This result as well as connection between  $S$-matrix
theory and approach with finite time interval for the diagrams of various type
will be presented in the next paper. We will also considered the another
exactly solvable problems and show that all the results are true for any
nuclear model.

In conclusion, we perform the direct calculation of the process amplitude with
annihilation products in the final state. Potential model is inapplicable to
the problem under study. This explains the different functional structure of
the results: $W(T)/W_{pot}(T)\sim \Gamma T$. If it is remembered that $T\sim
10^{31}y$ the quantitative distinction becomes clear as well.

\end{document}